\documentclass[floatfix,preprint]{revtex4}
\usepackage{amsmath,amssymb}
\usepackage[dvips]{graphicx}
\usepackage[dvips]{color}

\usepackage{mathrsfs}

\newcommand{\be}{\begin{equation}}
\newcommand{\ee}{\end{equation}}
\newcommand{\nablab}{\boldsymbol{\nabla}}
\newcommand{\IF}{\mathscr{I}}

\newcommand{\Tu}{T_{\!\boldsymbol{u}}}
\newcommand{\uu}{\boldsymbol{u}}

\newcommand{\rr}{\boldsymbol{r}}

\newcommand{\Pu}{\boldsymbol{P}_{\!\!\boldsymbol{u}}}

\newcommand{\PCN}{\boldsymbol{P}_{\!C,N}}

\newcommand{\ro}{\rho}
\newcommand{\p}{p}
\newcommand{\pn}{p_{\scriptscriptstyle N}}

\newcommand{\eq}[1]{Eq.~(\ref{#1})}

\begin{document}

\title{Fisher Information and Kinetic-energy Functionals: A Dequantization Approach}

\author{I. P. Hamilton}
\email{ihamilton@wlu.ca}\affiliation{Department of Chemistry,
Wilfrid Laurier University, Waterloo, Canada N2L 3C5.}

\author{Ricardo A. Mosna}
\email{mosna@ime.unicamp.br} \affiliation{ Instituto de
Matem\'atica, Estat\'\i stica e Computa\c{c}\~ao Cient\'\i fica,
Universidade Estadual de Campinas, C.P. 6065, 13083-859, Campinas,
SP, Brazil.}

\date{\today}

\begin{abstract}

We strengthen the connection between Information Theory and
quantum-mechanical systems using a recently developed dequantization
procedure whereby quantum fluctuations latent in the quantum
momentum are suppressed. The dequantization procedure results in a
decomposition of the quantum kinetic energy as the sum of a
classical term and a purely quantum term. The purely quantum term,
which results from the quantum fluctuations, is essentially
identical to the Fisher information. The classical term is
complementary to the Fisher information and, in this sense, it plays
a role analogous to that of the Shannon entropy. We demonstrate the
kinetic energy decomposition for both stationary and nonstationary
states and employ it to shed light on the nature of kinetic-energy
functionals.


\end{abstract}

\maketitle

\section{Introduction}
\label{sec:intro}

Over the past few years, Dehesa has been a pioneer in developing a
connection between information theory and quantum-mechanical
systems. \cite{dehesa5,dehesa6,dehesa1,dehesa2,dehesa3,dehesa4} This
connection is potentially of significant practical value as it is
related to density functional theory and, in particular, to the
construction of kinetic-energy functionals. For one-electron systems
with a central potential, such as the hydrogen atom, Dehesa obtained
an analytic expression for the Fisher information in terms of the
quantum numbers of the stationary states. \cite{dehesa2} Also over
the past few years we have developed a dequantization procedure,
first based on Witten deformation \cite{MHD05} and subsequently
based on a variational principle. \cite{MHD06} This dequantization
procedure results in a decomposition of the quantum kinetic energy
as the sum of a classical term and a purely quantum term. We
recently demonstrated the kinetic energy decomposition for
hydrogenic orbitals. \cite{HMD07}

In the present paper we examine connections between our work and
that of Dehesa. We consider our kinetic energy decomposition for
stationary states of the hydrogen atom and nonstationary states of a
particle in a box and a free particle represented as a Gaussian
wavepacket.

\subsection{Fisher Information}

The Fisher information, \cite{fisher,nagy} which is one of the
cornerstones of information theory, was developed as a measure of
spatial localization. For an $N$-electron system the Fisher
information is given by
\begin{equation}
\IF= \int \frac{|\nablab\p(\rr)|^2}{\p(\rr)} d^3\rr, \label{FI}
\end{equation}
where $\p(\rr_1)=\int|\psi(\rr_1,\ldots,\rr_N)|^2 \, d^3\rr_2\ldots
d^3\rr_N$ is the one-electron (probability) density. The electron
density, $\rho(\rr)$, is related to the one-electron (probability)
density by $\rho(\rr)=N\p(\rr)$. Thus the $\IF$ is a functional of
the electron density and is a local measure of the breadth of the
electron density. The greater the \emph{localization} of $\rho(\rr)$
the greater the value of the Fisher information. In contrast, the
greater the \emph{delocalization} of $\rho(\rr)$ the greater the
value of the Shannon entropy. \cite{Shannon} Thus the Fisher
information and the Shannon entropy are complementary quantities and
have been used in conjunction to analyze electron correlation and
other atomic properties. \cite{RD,SAA}

\subsection{Density Functional Theory}

Density functional theory (DFT) has developed into an extremely
successful approach for the calculation of atomic and molecular
properties. \cite{YangParr,DG,KH} In DFT, the electron density is
the fundamental variable and properties such as the energy are
obtained as a functional of $\rho(\rr)$ rather than from the
$N$-electron wavefunction, $\psi(\rr_1,\ldots,\rr_N)$ thereby
reducing a 3$N$-dimensional computation to a 3-dimensional one. The
energy can be partitioned into kinetic and potential terms and a
clear zeroth-order choice of functional for the potential energy is
the classical expression $-Ze^2\int \frac {\ro(\rr)}{r}d^3\rr +
\frac{e^2}{2}\int \int \frac
{\ro(\rr_1)\ro(\rr_2)}{r_{12}}d^3\rr_1d^3\rr_2$. However, for atomic
and molecular systems, there is no correspondingly clear
zeroth-order choice of functional for the kinetic energy.

A well-known kinetic-energy functional, formulated by Weizs\"{a}cker
\cite{weizsacker}, is

\begin{equation}
T_{W}=\frac{\hbar^2}{8m}\int\frac{|\nablab\ro(\rr)|^2}{\ro(\rr)}d^3\rr.
\label{TW}
\end{equation}
This expression is exact for the ground state of the hydrogen atom
(a one-electron system) but not for the ground states of
multi-electron atoms. Comparison of \eq{FI} and \eq{TW} shows that
the information content of the Fisher information and the
Weizs\"{a}cker term is the same and these quantities are essentially
identical (with $T_{W}=\frac{N\hbar^2}{8m}\IF$). In this paper we
generally employ the Weizs\"{a}cker term as the connection to the
kinetic energy is more direct.

\section{Background}

One of the key aspects of quantum mechanics is that one cannot
simultaneously ascribe well-defined (sharp) values for the position
and momentum of a physical system. This characteristic of quantum
mechanics is quantified by the position-momentum uncertainty
principle. \cite{heisenberg,kennard} We note that qualitatively
different position-momentum uncertainty relations based on the
Fisher information have recently been proposed.
\cite{dehesa2,dehesa6,dehesa4,dehesa1} Motivated by the
position-momentum uncertainty principle, quantization procedures
have been proposed in which the quantum regime is obtained from the
classical regime by adding a stochastic term to the classical
equations of motion. In particular, Nelson \cite{nelson,nelsonb} and
earlier work of F\'enyes \cite{FE} and Weizel \cite{WE} has shown
that the Schr\"odinger equation can be derived from Newtonian
mechanics via the assumption that particles are subjected to
Brownian motion with a real diffusion coefficient. The Brownian
motion results in an osmotic momentum and adding this term to the
classical momentum results in the quantum momentum.

\subsection{Dequantization}

We recently \cite{MHD06} proposed a dequantization procedure whereby
the classical regime is obtained from the quantum regime by
stripping these ``quantum fluctuations'' from the quantum momentum
resulting in the classical momentum. In particular, we introduced a
deformed momentum operator, which corresponds to generic
fluctuations of the particle's momentum. This leads to a deformed
kinetic energy which possesses a unique minimum that is seen to be
the classical kinetic energy. In this way, a variational procedure
determines the particular deformation that has the effect of
suppressing the quantum fluctuations, resulting in dequantization of
the quantum-mechanical system.


\section{Quantum-classical correspondence}
\label{sec:dequant}

For an $N$-electron system, consider a local deformation
$\boldsymbol{P}\to\Pu$ of the quantum momentum operator
$\boldsymbol{P}=-i\hbar\nablab$, with
\begin{equation}
\Pu\psi = \left( \boldsymbol{P} - i\uu \right) \psi,
\label{Pu}
\end{equation}
where all quantities in bold face are 3$N$-dimensional vectors and
$\uu$ is real.

Let
\begin{equation}
T=\frac{1}{2m}\int (\boldsymbol{P}\psi)^\ast (\boldsymbol{P}\psi) d^{3N}\rr \label{Tqm}
\end{equation}
and
\begin{equation}
\Tu =\frac{1}{2m}\int (\Pu\psi)^\ast (\Pu\psi) d^{3N}\rr \label{Th0}
\end{equation}
be the kinetic terms arising from $\boldsymbol{P}$ and $\Pu$,
respectively.

We recently \cite{MHD06} showed that extremization of $\Tu$ with
respect to $\boldsymbol{u}$-variations leads to the critical
$\boldsymbol{u}$ value
\begin{equation}
\uu_c = -\frac{\hbar}{2}\frac{\nablab\pn}{\pn}, \label{umin}
\end{equation}
where $\pn(\rr_1,\ldots,\rr_N)=|\psi(\rr_1,\ldots,\rr_N)|^2$ is the
$N$-electron (probability) density (with $\int \pn d^3\rr_1\cdots
d^3\rr_N$ = 1). This critical $\boldsymbol{u}$ value of results in
the $N$-electron classical momentum operator
\begin{equation}
\PCN \psi = \left( \boldsymbol{P} +
\frac{i\hbar}{2}\frac{\nablab\pn}{\pn} \right) \psi. \label{Pc}
\end{equation}
Thus our dequantization procedure \emph{automatically} identifies
the expression for $\uu_c$ which when added to the quantum momentum
results in the classical momentum. Here $-\uu_c$ is identical to the
osmotic momentum of Nelson \cite{nelson,nelsonb}, and adding
$-\uu_c$ to the classical momentum results in the quantum momentum.

This value of $\uu_c$ results in \be
T_{\uu_c}=T-\frac{\hbar^2}{8m}\IF_N = T - T_{W,N}, \label{Tumin} \ee
where $\IF_N$ is the $N$-electron Fisher information \be \IF_N= \int
\frac{\left( \nablab\pn \right)^2}{\pn} d^{3N}\rr \label{fisher} \ee
and $T_{W,N}$ is the $N$-electron Weizs\"{a}cker term.

If the wavefunction is written as $\psi=\sqrt{\pn}e^{iS_N/\hbar}$
where $S_N(\rr_1,\ldots,\rr_N)$ is the $N$-electron phase then a
straightforward calculation shows that the action of $\PCN$ on
$\psi$ is given by
\begin{equation}
\PCN\psi=\nablab S_N \: \psi, \label{Pcl1}
\end{equation}
so that, from \eq{Th0},
\begin{equation}
T_{\uu_c}=\frac{1}{2m} \int \pn \, |\nablab S_N|^2 d^{3N}\rr.
\label{Tuc2}
\end{equation}
This quantity is the mean kinetic energy of a classical ensemble,
described by the density $\pn$ and momentum $\nablab S_N$
\cite{Goldstein,Holland} and we therefore refer to $T_{\uu_c}$ as
the $N$-electron classical kinetic energy $T_{C,N}$.


\section{Results and Discussion}
\label{sec:decomp}

From \eq{Tumin}, the $N$-electron kinetic energy can be expressed as
\begin{equation}
T_N = T_{C,N} + T_{W,N}. \label{dec_T0}
\end{equation}
This is the sum of the $N$-electron classical kinetic energy and the
$N$-electron Weizs\"{a}cker term which is purely quantum and results
from the quantum fluctuations. We showed \cite{HMD07} previously
that the $N$-electron Weizs\"{a}cker term can be decomposed as $T_W$
(a one-electron term) and a purely quantum kinetic correlation term,
$T_Q^{corr}$. Furthermore we showed that $T_W$ results from the
local quantum fluctuations while $T_Q^{corr}$ results from the
nonlocal quantum fluctuations. Then, assuming that the $N$-electron
classical kinetic energy can be decomposed as $T_{C}$ (a
one-electron term) and a classical kinetic correlation term,
$T_{C}^{corr}$, we can write

\begin{equation} T_N = T_{C} + T_{C}^{corr} + T_W + T_{Q}^{corr}. \label{dec_Tone}
\end{equation}


\subsection{Noninteracting kinetic energy}
\label{sec:noni}

In the orbital approximation, kinetic correlation is neglected.
Omitting these terms in Eq.~(\ref{dec_Tone}), we obtain the
noninteracting kinetic energy as
\begin{equation}
T_s = T_{C}+ T_W. \label{dec_Ts}
\end{equation}
There are two limiting cases for which this expression can be
obtained analytically. For the ground state of the hydrogen atom (an
$N$ = 1 system), the electron phase is zero, so $T_{C}=0$.
Therefore, $T_s = T_W$ which is the correct result for this limit.
For the uniform electron gas (an $N = \infty$ system) $\ro$ is
uniform so $T_W=0$. Therefore $T_s = T_{C}$ which can be calculated
by adding up the kinetic energies of one-electron orbitals
approximated as local plane waves. This results in the Thomas-Fermi
term \cite{thomas,fermi,yang} which is the correct result for this
limit.

\subsection{One-particle systems}

For a one-particle system the noninteracting kinetic energy is
simply the kinetic energy and Eq.~(\ref{dec_Ts}) becomes $T = T_{C}
+ T_W$. We note that the integrands of $T_{C}$ and $T_W$
($\mathcal{T}_C$ and $\mathcal{T}_W$) are never negative and
correspondingly, $T_{C}$ and $T_W$ are never negative. Thus both
$T_{C}$ and $T_W$ are lower bounds to the kinetic energy. In the
next two subsections we explicitly show that our expression for the
kinetic energy is correct for both stationary and nonstationary
states. Furthermore:
\begin{equation}
\mathcal{T} = \mathcal{T}_{C} + \mathcal{T}_W. \label{dec_dT}
\end{equation}
\noindent That is, the integrand of $T$ is equal to the sum of the
integrands of $T_{C}$ and $T_W$. This is the case for all values of
the position at each value of the time.

\subsection{Stationary states}

The hydrogenic orbitals, $\psi{(n,l,m)}$, are dependent on the
principal quantum number $n$, the angular momentum quantum number
$l$ and the magnetic quantum number $m$ but the total energy is
dependent only on $n$ and is (in atomic units) $E$ = -1/2$n^2$.
Then, from the virial expression for Coulombic systems, the kinetic
energy is $T$ = -$E$ = 1/2$n^2$. For $n$ = 1, the classical kinetic
energy is zero.

We previously \cite{HMD07} presented results for $n$ = 2 which is
the first nontrivial case and here we present results for $n$ = 3.
The classical kinetic energy is zero for $\psi{(3,0,0)}$,
$\psi{(3,1,0)}$ and $\psi{(3,1,0)}$ and, from direct calculation,
$T_W$ = 1/18 which is equal to $T$. However, the classical kinetic
energy is nonzero for $\psi{(3,1,1)}$ and $\psi{(3,1,-1)}$ and, from
direct calculation, $T_{C}$ = 1/54 and $T_W$ = 1/27 and $T_{C}+ T_W$
= 1/18 which is equal to $T$. Radial distributions (integrated over
the angular variables) of the integrands for $T_{C}$, $T_W$ and $T$
for this case are shown in Fig. 1(a).

The radial distribution for $T_{C}$ is dependent on $n$, $l$ and
$|m|$ but the classical kinetic energy is dependent only on $n$ and
$|m|$ and $T_{C}$ = $\frac {|m|}{n}T = |m|/2n^3$. Correspondingly,
$T_W$ = $\frac {n - |m|}{n}T = (n-|m|)/2n^3$ and we note that this
expression could be deduced from the analytic expression for the
Fisher information obtained by Dehesa. \cite{dehesa2} We also note
that whereas $T_{C}$ can equal zero, $T_W$ cannot since, for a
normalizable state, $\rho$ cannot be uniformly constant and
therefore $\nabla\rho$ cannot be identically zero. The fact that the
purely quantum term cannot equal zero is in complete accord with the
position-momentum uncertainty principle. From the above expressions,
$T_{C}$ and $T_W$ are constant for $n$ and $|m|$ fixed and this is
illustrated in Fig. 1(b) which shows the radial distributions for
$T_{C}$, $T_W$ and $T$ for $n$ = 3, $l$ = 2 and $|m|$ = 1. Although
the radial distribution for $T_{C}$ and $T_W$ are clearly different
from those of Fig. 1(a) they again integrate to 1/54 and 1/27
respectively and $T_{C}$ + $T_W$ = 1/18. For $n$ and $l$ fixed,
$T_{C}$ increases from 0 to $l/2n^3$ while $T_W$ decreases from
$1/2n^2$ to $(n-l)/2n^3$ as $|m|$ increases from 0 to $l$ and this
is illustrated in Fig. 1(c) which shows the radial distributions for
$T_{C}$, $T_W$ and $T$ for $n$ = 3, $l$ = 2 and $|m|$ = 2. In this
case the radial distributions for $T_{C}$ and $T_W$ integrate to
1/27 and 1/54 respectively and we again have $T_{C}$ + $T_W$ = 1/18.
The results for these stationary states support our expression for
the kinetic energy . Furthermore, it is clear from Fig. 1 that the
integrand of $T$ is equal to the sum of the integrands of $T_{C}$
and $T_W$ for all values of the position.

\begin{figure*}[htbp]
\begin{center}
\includegraphics[width=6cm]{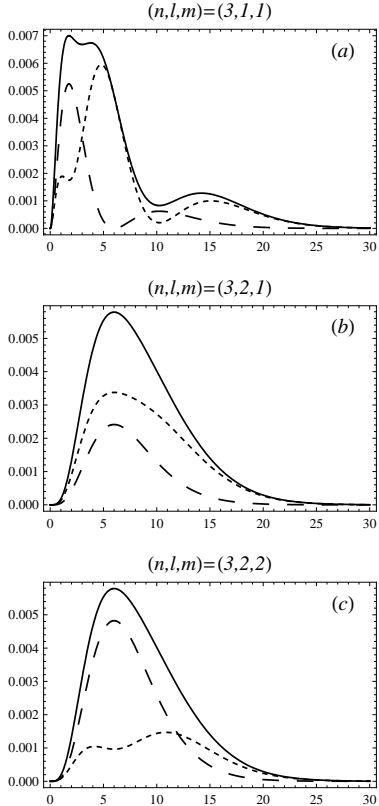}
\caption{Radial distributions (integrated over the angular
variables) of $\mathcal{T}_{C}$ (dashed curve), $\mathcal{T}_W$
(dotted curve) and $\mathcal{T}$ (solid curve) for hydrogenic
orbitals with $n$ = 3 and (a) $l$ = 1, $|m|$ = 1; (b) $l$ = 2,
$|m|$; (c) $l$ = 2, $|m|$ = 2. The horizontal axis is in atomic
units.} \label{fig:fig1}
\end{center}
\end{figure*}

\vfil\eject

\subsection{Nonstationary states}

We first consider a one-dimensional particle in a box (pib) state
that is initially $\phi(x) = 2^{-1/2}(\psi_1(x)+\psi_2(x))$ where
$\psi_1(x)$ and $\psi_2(x)$ are the first two pib eigenfunctions.
The pib eigenfunctions are, of course, stationary states and, for
both $\psi_1(x)$ and $\psi_2(x)$, $T_{C}$ = 0 and $T = T_W$.
However, $\phi(x)$ is a nonstationary state and whereas $T_{C}$ = 0
for $t$ = 0, this is generally not the case for later times. This is
clear from Fig. 2 which shows the probability distribution (upper
panel) and integrands for $T_{C}$, $T_W$ and $T$ (lower panel). Note
that at $t$ = 0.075 there is a relatively flat shoulder on the right
side of the probability distribution and that in this region,
$\mathcal{T}_W$ is small whereas $\mathcal{T}_C$ is large. At $t$ =
0.150 there is a relatively flat shoulder on the left side of the
probability distribution for which this is also the case.

We now consider a free particle represented as a one-dimensional
Gaussian wavepacket that is initially $\phi(x)= \pi^{-1/4}e^{-x^2
/2}$. Again, $T_{C}$ = 0 for $t$ = 0 but it is clear from Fig. 3,
which shows the probability distribution (upper panel) and
integrands for $T_{C}$, $T_W$ and $T$ (lower panel), that, as the
Gaussian wavepacket spreads, $T_{C}$ increases while $T_W$ decreases
and that as $t$ $\rightarrow$ $\infty$, $T_{C}$ $\rightarrow$ $T$
while $T_W$ $\rightarrow$ 0. Note that as $t$ increases and the
probability distribution becomes flatter, $\mathcal{T}_W$ becomes
smaller whereas $\mathcal{T}_C$ becomes larger. Thus as the particle
becomes delocalized there is a transition from purely quantum
kinetic energy to classical kinetic energy.

The results for these nonstationary states support our expression
for the kinetic energy. Furthermore, it is clear from Figs. 2 and 3
that the integrand of $T$ is equal to the sum of the integrands of
$T_{C}$ and $T_W$ for all values of the position at each value of
the time.

\begin{figure*}[htbp]
\begin{center}
\includegraphics[width=0.7\linewidth]{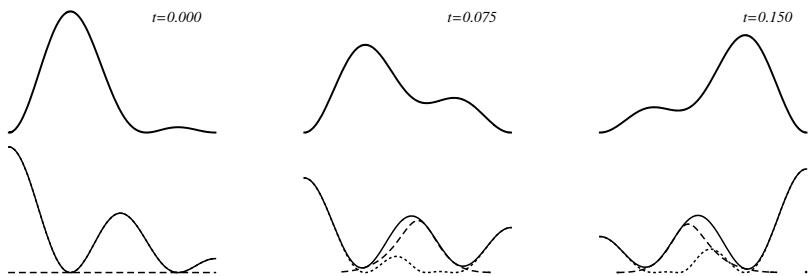}
\caption{One-dimensional pib state that is initially $\phi(x) =
2^{-1/2}(\psi_1(x)+\psi_2(x))$ where $\psi_1(x)$ and $\psi_2(x)$ are
the first two pib eigenfunctions. Distributions shown at $t$ =
0.000; $t$ = 0.075; $t$ = 0.150: Probability distribution (upper
panel); Distributions of $\mathcal{T}_{C}$ (dashed curve),
$\mathcal{T}_W$ (dotted curve) and $\mathcal{T}$ (solid curve).
(lower panel)} \label{fig:fig2}
\end{center}
\end{figure*}

\begin{figure*}[htbp]
\begin{center}
\includegraphics[width=0.7\linewidth]{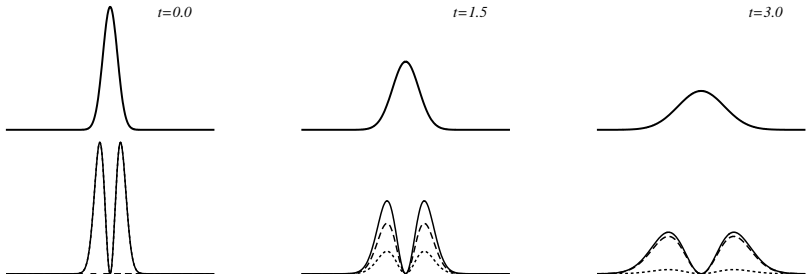}
\caption{One-dimensional Gaussian wavepacket that is initially
$\phi(x)= \pi^{-1/4}e^{-x^2 /2}$ but for which there is no confining
potential. Distributions shown at $t$ = 0.0; (b) $t$ = 1.5; (c) $t$
= 3.0: Probability distribution (upper panel); Distributions of
$\mathcal{T}_{C}$ (dashed curve), $\mathcal{T}_W$ (dotted curve) and
$\mathcal{T}$ (solid curve) (lower panel).} \label{fig:fig3}
\end{center}
\end{figure*}

\vfil\eject

\section{Conclusions}

In Nelson's quantization procedure an osmotic momentum term is added
to the classical momentum resulting in the quantum momentum. This
osmotic momentum term represents the quantum fluctuations that are
an essential part of quantum mechanics in accord with the
position-momentum uncertainty principle. In our dequantization
procedure this osmotic momentum term is removed from the quantum
momentum resulting in the classical momentum. We obtain the osmotic
momentum term via a variational approach in which the deformed
quantum kinetic energy is minimized with respect to variations of
the deformation parameter. The critical value of the deformation
parameter which minimizes the deformed kinetic energy is directly
related to the osmotic momentum term.

The result of our dequantization procedure is the decomposition of
the kinetic energy into the classical kinetic energy and the purely
quantum kinetic energy. The purely quantum kinetic energy is the
Weizs\"{a}cker term which is essentially identical to the Fisher
information. The purely quantum kinetic energy is thereby a direct
functional of the electron density and is a critical component of
the kinetic-energy functional (and this is well-known). However, the
classical kinetic energy is also a critical component of the
kinetic-energy functional. Unfortunately, the classical kinetic
energy is explicitly dependent on the phase of the wavefunction and
is manifestly not a direct functional of the electron density.
Devising a functional of the electron density that indirectly but
accurately approximates classical kinetic energy is a major
challenge for the development of quantitative kinetic-energy
functionals.

It is well-known that the Weizs\"{a}cker term, $T_W$, which is
greater than or equal to zero, is a lower bound to the kinetic
energy, $T$. We have shown that the classical kinetic energy,
$T_{C}$, which is also greater than or equal to zero, is also a
lower bound to the kinetic energy and $T$ = $T_W$ + $T_{C}$.
Furthermore, we have shown that the integrands of the Weizs\"{a}cker
term and the classical kinetic energy (which are both greater than
or equal to zero) are each lower bounds to the integrand of the
kinetic energy and $\mathcal{T}$ = $\mathcal{T}_W$ +
$\mathcal{T}_{C}$. Examples have been given for which this is the
case at each value of the position and, for nonstationary states,
for which this is also the case at each value of the time. It is
well-established that the Fisher information and the Shannon entropy
are complementary quantities. We have shown that the Weizs\"{a}cker
term (which is essentially identical to the Fisher information) and
the classical kinetic energy are complementary quantities and, in
this sense, the classical kinetic energy is analogous to the Shannon
entropy.

\acknowledgments

IPH acknowledges funding from NSERC and RAM acknowledges FAPESP for
financial support.

\end{document}